\documentclass[12pt]{article}
\usepackage{color}
\usepackage[dvips]{graphicx}
\usepackage{epsfig,cite}
\usepackage {amsmath}
\usepackage{rotating}
\usepackage{amssymb}
\usepackage{slashed}
\include{epsf}
\newcount\timecount
\newcount\hours \newcount\minutes  \newcount\temp \newcount\pmhours
\hours = \time
\divide\hours by 60
\temp = \hours
\multiply\temp by 60
\minutes = \time
\advance\minutes by -\temp
\def\hour{\the\hours}
\def\minute{\ifnum\minutes<10 0\the\minutes
            \else\the\minutes\fi}
\def\clock{
\ifnum\hours=0 12:\minute\ AM
\else\ifnum\hours<12 \hour:\minute\ AM
      \else\ifnum\hours=12 12:\minute\ PM
            \else\ifnum\hours>12
                 \pmhours=\hours
                 \advance\pmhours by -12
                 \the\pmhours:\minute\ PM
                 \fi
            \fi
      \fi
\fi
}

\def\monthname{\relax\ifcase\month 0/\or January\or February\or
   March\or April\or May\or June\or July\or August\or September\or
   October\or November\or December\else\number\month/\fi}

\def\bold#1{\setbox0=\hbox{$#1$}%
     \kern-.025em\copy0\kern-\wd0
     \kern.05em\copy0\kern-\wd0
     \kern-.025em\raise.0433em\box0 }

\def\URLtilde{\lower0.2em\hbox{$\tilde{\phantom{a}}$}}
\def\mycomm#1{\hfill\break\strut\kern-3em{\color{red}\tt ====> #1\color{black}}\hfill\break}

\def\beq{\begin{equation}}
\def\eeq{\end{equation}}

\def\ga{\mathrel{\raise.3ex\hbox{$>$\kern-.75em\lower1ex\hbox{$\sim$}}}}
\def\la{\mathrel{\raise.3ex\hbox{$<$\kern-.75em\lower1ex\hbox{$\sim$}}}}
\def\gev{{\rm \, Ge\kern-0.125em V}}
\def\tev{{\rm \, Te\kern-0.125em V}}
\def\gyr{{\rm \, G\kern-0.125em yr}}

\def\gappeq{\mathrel{\rlap {\raise.5ex\hbox{$>$}}
{\lower.5ex\hbox{$\sim$}}}}
\def\lappeq{\mathrel{\rlap{\raise.5ex\hbox{$<$}}
{\lower.5ex\hbox{$\sim$}}}}
\def\Toprel#1\over#2{\mathrel{\mathop{#2}\limits^{#1}}}

\def\m12{m_{1\!/2}}

\def\bea{\begin{eqnarray}}
\def\eea{\end{eqnarray}}

\def\pref#1{(\ref{#1})}

\def\beqar{\begin{eqnarray}}
\def\eeqar{\end{eqnarray}}

\def\m{{\cal m}}

\newcommand{\lsim}{\mathrel{\hbox{\rlap{\lower.55ex
\hbox{$\sim$}} \kern-.3em \raise.4ex \hbox{$<$}}}}
\newcommand{\gsim}{\mathrel{\hbox{\rlap{\lower.55ex
\hbox{$\sim$}} \kern-.3em \raise.4ex \hbox{$>$}}}}

\begin{document}
\begin{titlepage}
\pagestyle{empty}
\baselineskip=21pt
\rightline{KCL-PH-TH/2012-18, LCTS/2012-10, CERN-PH-TH/2012-103, TAUP-2949/12}
\vskip 1in
\begin{center}
{\large {\bf Indications on the Mass of the Lightest Electroweak Baryon}}
\end{center}
\begin{center}
\vskip 0.5in
 {\bf John~Ellis}$^{1,2}$ and {\bf Marek~Karliner}$^3$
\vskip 0.3in
{\small {\it
$^1${Theoretical Particle Physics and Cosmology Group, Physics Department, \\
King's College London, London WC2R 2LS, UK}\\

$^2${TH Division, Physics Department, CERN, CH-1211 Geneva 23, Switzerland}\\

$^3${Raymond and Beverly Sackler School of Physics and Astronomy,
Tel Aviv University, Israel}\\

}}
\vskip 0.5in
{\bf Abstract}\\

\end{center}
In general, an effective low-energy Lagrangian model of composite electroweak
symmetry breaking contains soliton solutions that may be identified with technibaryons.
We recall how the masses of such states may be related to the coefficients of
fourth-order terms in the effective Lagrangian, and review the qualitative success of this approach
for baryons in QCD. We then show how the current theoretical and phenomenological constraints on the
corresponding fourth-order coefficients in the electroweak theory could be used
to estimate qualitative lower and upper bounds on the lightest electroweak baryon mass. We also discuss
how the sensitivity of the LHC experiments could enable these bounds to be improved. 

\baselineskip=18pt \noindent
{\small
}
\vfill
\leftline{
April 2012}
\end{titlepage}
\baselineskip=18pt

\section{Introduction}

One of the primary objectives of the LHC experimental programme is to explore the
electroweak symmetry-breaking sector. Theoretical perspectives range between the possibility that
there is a single elementary Higgs-like boson as in the Standard Model~\cite{Higgs}, and the possibility that electroweak symmetry is
broken dynamically by a new strongly-interacting sector that might not yield a recognizable scalar boson, as in
traditional technicolour models~\cite{TC}. Intermediate possibilities include composite scenarios in which there is a relatively light
pseudo-Nambu-Goldstone boson whose couplings might differ observably from those of a
Standard Model Higgs boson~\cite{PNGB}.

The LHC experiments ATLAS~\cite{ATLAS} and CMS~\cite{CMS} currently exclude a Standard Model-like Higgs boson weighing
$< 122.5$~GeV and between 127.5~GeV and 600~GeV. Open possibilities include a heavy Higgs-like boson
as in traditional composite models, a (possibly elementary) Higgs-like boson with mass $125 \pm 2.5$~GeV,
and an intermediate-mass scalar boson with a smaller production cross section and/or observable decay branching ratios than
a Standard Model Higgs boson. Present LHC data hint at the existence of a scalar boson weighing $\sim 125$~GeV~\cite{ATLAS,CMS},
but this remains to be confirmed.

Independently from the possible existence of a recognizable scalar boson, there are already indications
that any strongly-interacting electroweak symmetry-breaking sector could not simply be a scaled-up
version of QCD. For example, QCD-like models would make excessive contributions to the
electroweak vacuum polarization $S$~\cite{bigS}, and potentially also to $T$~\cite{bigT}. For this reason, there is much
discussion of `walking technicolour' models~\cite{WTC} in which the coupling strength of the new strong interaction
evolves relatively slowly over an extended range of energy scales, reflecting nearly-conformal dynamics. 
Such a scenario may also help suppress flavour-changing neutral interactions in extended technicolour models. 

It is interesting to glean as much information as possible about the possibilities allowed for any such 
novel strongly-interacting sector. For example, in some such scenarios the effective low-energy
spectrum contains a pseudo-dilaton~\cite{pseudoD,CEO3}, a pseudo-Nambu-Goldstone boson of approximate conformal
invariance. The above-mentioned LHC data constrain such a possibility over a wide range of possible
pseudo-dilaton masses~\cite{CEO3,boundD}. A complementary window on possible strongly-interacting electroweak symmetry-breaking
sectors could be provided by technibaryons, the lightest of which would be stable in many models and
therefore could be a candidate for cold dark matter~\cite{Shmuel}.

Another window on the electroweak symmetry-breaking sector is provided by the search for
non-standard interactions between massive vector bosons $V$, either indirectly via their effects
on precision electroweak observables or directly via measurements of $VV$ scattering at the LHC.
Lower limits on $VV$ interactions of fourth order in particle momenta (field derivatives) have been derived
from crossing, Lorentz invariance and unitarity~\cite{DGPR}, which are considerably smaller than the 
upper limits that have been derived from precision electroweak data~\cite{EGM}. It has also been
argued that the LHC experiments should have the sensitivity to improve significantly on the current
upper limits~\cite{EGM}.

The purpose of this paper is to connect the technibaryon and $VV$ scattering windows on a possible
strongly-interacting electroweak sector, using the
Skyrme model of baryons~\cite{Skyrme:1961vq,Adkins:1983ya}. The second-order terms
in the effective Lagrangian for electroweak symmetry breaking are isomorphic to the
corresponding terms in the low-energy SU(2) $\times$ SU(2) chiral Lagrangian for QCD, which
is known, in the presence of suitable generic fourth-order terms,  to possess `Skyrmion' soliton 
solutions that can be identified  quite successfully with baryons~\cite{SkyrmeOK,Weigel}. It is therefore natural also to use 
the Skyrme model to describe technibaryons, at least qualitatively~\cite{CEO3}. The properties of Skyrmions,
in particular their masses, are related to the magnitudes of the fourth-order terms in the
effective Lagrangian. Therefore, constraints on (measurements of) anomalous $VV$
scattering can be used to bound (estimate) the mass of the lightest technibaryon, a quantity of
potential interest to astrophysical searches for dark matter as well as the LHC experiments.

We find that the present constraints on anomalous $VV$ scattering do not provide very strong
constraints on the lightest technibaryon mass, though they might be of some interest for some
walking technicolour models. On the other hand, LHC measurements of $VV$ scattering
could provide much more restrictive information on the possible appearance of technibaryons
in the multi-TeV range.

\section{Phenomenological Framework}

We work within the framework of the following nonlinear low-energy effective Lagrangian for the
electroweak symmetry-breaking sector:
\bea
{\cal L}_{\it eff} \; & = & \; \frac{v^2}{4} {\rm Tr} \left(D_\mu U D^\mu U^\dagger \right) + \frac{1}{32e^{2}} S + \dots ,
\label{effL}
\eea
where $U$ is a unitary $2 \times 2$ matrix parametrizing the three Nambu-Goldstone fields
that are `eaten' by the $W^\pm$ and $Z^0$, giving them masses, $v \sim 246$~GeV is the
conventional electroweak scale, and the second, 
`Skyrme term'~\cite{Skyrme:1961vq,Adkins:1983ya}
\beq
S \; \equiv \; {\rm Tr}[(D_\mu U)U^\dagger,(D_\nu U)U^\dagger]^{2} 
\label{Skyrme}
\eeq
is scaled by an {\it a priori} unknown parameter $e$. The $\dots$ represent other possible terms of
fourth and higher orders in a derivative expansion.
The effective low-energy chiral Lagrangian for QCD has the same form as (\ref{effL}), with
the electroweak scale $v$ replaced by the pion decay constant $f_\pi \sim 93$~MeV.

The effective Lagrangian (\ref{effL}) necessarily has soliton solutions, because $\pi_3$(SU(2)) = Z,
which have integer baryon number
\begin{equation}
B = \frac{1}{24 \pi^2} \int d^3x \epsilon^{ijk} {\rm Tr}\left[(U^{-1}\partial_iU)(U^{-1}\partial_jU)(U^{-1}\partial_kU)\right].
\label{Bnumber}
\end{equation}
In the case of a nonlinear SU(3) $\times$ SU(3) $\to$ SU(3) theory
the effective Lagrangian contains
a Wess-Zumino term $N \Gamma$ \cite{Wess:1971yu}, 
where $N$ is an integer, and the lowest-lying
$B = 1$ baryon is a fermion (boson) with $I = J = {1\over2}$ $(I = J = 0)$ if $N$ is odd (even). If the
underlying strongly-interacting theory is a non-Abelian SU(N) gauge theory with fermions, $N$ is
identified with the number of colours, and in QCD the $B = 1$ baryon is necessarily a fermion because $N = 3$. 
In the case of the nonlinear SU(2) $\times$ SU(2) $\to$ SU(2) theory (\ref{effL}),
there is no Wess-Zumino term, and the $B = 1$ electroweak baryon may be either a boson or a fermion,
with the latter remaining a topological possibility, since $\pi_4$(SU(2)) =
Z$_2$~\cite{Finkelstein:1968hy,Witten:1983tx}.
Here we leave
open the question whether the electroweak baryonic `Skyrmions' are bosons or fermions. 

The masses and other properties of Skyrmions can be calculated in the semiclassical
limit of large $N$. We consider such calculations to be quite successful in QCD, where $N = 3$,
and assume that they are qualitatively correct also in the electroweak sector.
If the masses of the strongly-interacting fermions can be neglected, as assumed in
technicolour models, numerical calculation of the lightest Skyrmion mass 
yields~\cite{Adkins:1983ya}
\beq
M_S \; \sim \; \frac{73}{e} v + \dots ,
\label{MS}
\eeq
where the dots represent corrections due, e.g., to collective quantization of the
rotational degrees of freedom in the fermionic case. In the case of QCD, we replace $v$ by $f_\pi$,
and for the $I = J = {1\over2}$ nucleon
the rotational correction takes the form
\beq
\Delta M_S \; = \; \frac{3}{8 \lambda} ,
\label{MNcorrection}
\eeq
where $\lambda \sim  (2 \pi/3) (25.5/e^3 f_\pi)$, and the correction is a
factor 5 larger in the case of the $I = J = {3\over2}$ \;$\Delta$
multiplet.

The fourth-order terms in the effective Lagrangian (\ref{effL}) are often parametrized as follows:
\beq
{\cal L}_{\it eff4} \; = \; \alpha_4 \, {\rm Tr} \left[ \left(V_\mu V^\nu
\right)^2 \right] \; + \; \alpha_5 \, {\rm Tr} \left[ \left(V_\mu V^\mu \right)^2 \right] ,
\label{alpha4alpha5}
\eeq
where $V_\mu \equiv (D_\mu U)U^\dagger$. The Skyrme term (\ref{Skyrme}) is the antisymmetric combination
\linebreak
$S = 2 \{ {\rm Tr} \left[ \left(V_\mu V^\nu \right)^2 \right] - {\rm Tr} \left[ \left(V_\mu V^\mu \right)^2 \right] \}$,
and we introduce the orthogonal, symmetric combination 
$T \equiv 2 \{ {\rm Tr} \left[ \left(V_\mu V^\nu \right)^2 \right] + {\rm Tr} \left[ \left(V_\mu V^\mu \right)^2 \right] \}$.
\,Writing \,${\cal L}_{\it eff4} \equiv s \cdot S + t \cdot T$, we see that
\beq
s \; = \frac{1}{32 e^2} \; = \; \frac{\alpha_4 - \alpha_5}{4}
\raise0.2em\hbox{,}
\quad 
t \; = \;  \frac{\alpha_4 + \alpha_5}{4} 
\raise0.2em\hbox{.}
\label{st}
\eeq
and from eqs.~\pref{MS} and \pref{st}, using $v = 246$~GeV we find
\beq
M_S \sim 102 \sqrt s  \ \ {\rm TeV} .
\label{MSfroms}
\eeq
In the following we use these relations and
various theoretical and phenomenological bounds on $\alpha_{4,5}$ to constrain $e$ and
hence estimate the mass of the lightest electroweak technibaryon within the Skyrmion framework.

\section{Bounds on Fourth-Order Chiral Lagrangian Terms and the QCD Skyrmion Mass}

In QCD a fit to the $I = J = {1\over2}$\, nucleon and the $I = J =
{3\over2}$ \,$\Delta$ masses, assuming that their
mass difference of $\sim 300$~MeV is provided by the semiclassical collective quantization of the rotational
degree of freedom of the Skyrmion (\ref{MNcorrection}) and neglecting the light quark masses, 
yielded the value of 65~MeV for $f_\pi$ (to be compared with the
physical value $f_\pi = 93$~MeV) and $e = 5.65$~\cite{Adkins:1983ya}.
Bearing in mind the discrepancy in $f_\pi$, this qualitative historical success suggests that the uncertainty in the estimate of the 
Skyrmion mass is $\sim {\cal O}(30)$\%. 

Here we take a different approach, more similar to what we use later in the electroweak case,
taking $f_\pi = 93$~MeV from experiment and using data to estimate the magnitude of the
Skyrme term in QCD. A global fit to the fourth-order terms in the effective low-energy Lagrangian
of QCD~\cite{EGPR} yields~\cite{Pichfit}
\bea
\alpha_4 & = & \phantom{{-}} ( 1.4 \pm 0.3) \times 10^{-3}, \label{QCDvalue1} \\
\alpha_5 - \alpha_4 & = & ({-} 2.7 \pm 1.3) \times 10^{-3} ,
\label{QCDvalue2}
\eea
when these parameters are defined at a renormalization scale equal to $M_\rho$.
These ranges are in good qualitative agreement with the predictions of the $1/N_c$ expansion.
Inserting them into (\ref{st}) yields
\beq
s \; = \frac{1}{32 e^2} \; = \; (0.7 \pm 0.3) \times 10^{-3} ,
\label{svalue}
\eeq
and hence $e = 7^{+3}_{-1}$. 

This range lies somewhat above the value $e = 5.45$ found in~\cite{Adkins:1983ya} from a combined fit to the nucleon
and $\Delta$ masses, 
but we are encouraged that it is within a similar ballpark. Inserting this range into (\ref{MS}) 
and replacing $v$ by $f_\pi = 93$~MeV yields the estimate $\sim 1000$~MeV
for the nucleon mass, before incorporating the semiclassical correction (\ref{MNcorrection})
associated with collective quantization of the rotational degree of freedom. This is $\sim 220$~MeV for the
central value $e \sim 7$, but is $\propto e^3$ and hence quite uncertain. Nevertheless, we are
encouraged that the estimated nucleon mass is correct to within about 30\%, consistent with the
expected uncertainty.

\section{Bounds on Fourth-Order Electroweak Terms and the Techni-Skyrmion Mass}

In the case of the electroweak theory, lower bounds on the coefficients $\alpha_{4,5}$ were obtained in~\cite{DGPR}, 
assuming just crossing, Lorentz invariance and unitarity:
\bea
\alpha_4 (v) \; & > & \; 6 \times 10^{-4} , \label{Grinsteinbound1} \\
\alpha_4 (v) + \alpha_5 (v) \; & > & \; 1.1 \times 10^{-3} ,
\label{Grinsteinbound2}
\eea
where we have noted that $\alpha_{4,5}$ are specified at the scale of electroweak symmetry breaking~\cite{DGPR}.
As might have been expected from the generality of the input assumptions, it is
not possible to derive useful constraints on the technibaryon mass from
(\ref{Grinsteinbound2}). These constraints are clearly compatible with either
$s \to 0$, in which case (\ref{MSfroms}) would allow an arbitrarily small Skyrmion mass,
or $s \to \infty$, in which case $M_S$ would be arbitrarily large. That said, one might be tempted to
consider the possibility that both the bounds (\ref{Grinsteinbound1}, \ref{Grinsteinbound2}) were saturated,
with the inference that
\beq
s \; = \frac{1}{32 e^2} \; \sim \; 0.25 \times 10^{-4}, \; \; t \; \sim \; 2.8 \times 10^{-4} ,
\label{equality}
\eeq
corresponding naively to $e \sim 35$ and $M_S \sim 0.5$~TeV. However, such an
inference would be unsound, since $\alpha_{4,5}$ are subject to important renormalization effects~\cite{DGPR}:
\bea
\alpha_4 (\mu) \; & = & \; \alpha_4 (v) + \frac{1}{96 \pi^2} {\rm ln}\left( \frac{v}{\mu} \right) , \label{Grinsteinrenn1} \\
\alpha_5 (\mu) \; & = & \; \alpha_5 (v) + \frac{1}{192 \pi^2} {\rm ln}\left( \frac{v}{\mu} \right) .
\label{Grinsteinrenn2}
\eea
If one were to use $\mu \sim 500$~GeV, one would find a reduction in $s$ by $\sim 0.9 \times 10^{-4}$, 
preventing the derivation of any estimated bound on $M_S$.

Additional dynamical information is required if interesting bounds on $M_S$ are to be derived.
For example, upper and lower limits on the magnitudes of the coefficients
$\alpha_{4,5}$ were obtained in~\cite{EGM},
on the basis of precision electroweak data:
\bea
- 0.35 \; < & \alpha_4 (v) \; & <  \; 0.06 , \nonumber \\
-0.87 \; < & \alpha_5 (v) \; & <  \; 0.15 .
\label{Conchabounds}
\eea
These are compatible with $\alpha_4 = \alpha_5$ and hence $s = 0$, and so not provide
a lower bound on the electroweak baryon mass. However, an upper bound on $M_S$ can be obtained by substituting
$\alpha_4 < 0.06$ and $\alpha_5 >  - 0.87$ into (\ref{st}), obtaining $s <
0.23$, 
and hence
\beq
M_S \; < \;  50 \; {\rm TeV} .
\label{presentbound}
\eeq
This constraint is quite weak, though it might be relevant in the context of some models of
`walking technicolour'.

It was also estimated in~\cite{EGM} that non-observation of $\alpha_{4,5}$ by the LHC experiments
ATLAS and CMS would restrict these coefficients to the ranges
\bea
- 7.7 \times 10^{-3} \; < & \alpha_4 (v) \; & <  \; 15 \times 10^{-3} , \nonumber \\
- 12 \times 10^{-3} \; < & \alpha_5 (v) \; & <  \; 10 \times 10^{-3} .
\label{LHCsensitivity}
\eea
If $\alpha_{4,5}$ were indeed not to be measured at these levels, one could conclude that $s < 7 \times 10^{-3}$ and $e > 2.1$, 
 hence
 \beq
M_S \; \lsim \;  \; 8.5 \; {\rm TeV} .
\label{futurebound}
\eeq
We note that both the bounds (\ref{presentbound}, \ref{futurebound}) are reasonably insensitive
to the inclusion of the renormalization effects (\ref{Grinsteinrenn1}, \ref{Grinsteinrenn2}), which
would change $\alpha_{4,5}$ by $\sim 4 \times 10^{-3}, 2 \times 10^{-3}$ even if one chose $\mu = 10$~TeV.
The possible future constraint (\ref{futurebound}) is much stronger than the present bound (\ref{presentbound}),
and may be of relevance to a wider class of technicolour models.

We have neglected in this Section the semiclassical correction to the Skyrmion mass due to
collective quantization of the rotational degrees of freedom (\ref{MNcorrection}), that appears if the technibaryon is
a fermion but would be absent if it is a boson. This may be estimated on the basis of (\ref{MNcorrection}) to be
$\Delta M \approx 7 \times 10^{-3} e^3 v = 9.5/s^{3/2}$~MeV, corresponding to $\approx 16$~GeV
if the bound (\ref{LHCsensitivity}) is saturated. In fact, as argued in~\cite{CEO3}, 
there are at least two reasons why the technibaryon should be a boson, at least if it is stable on a cosmological
time scale. One is that the cold dark matter scattering cross section of a fermionic
technibaryon is likely to exceed the experimental upper limit~\cite{BDT}. The other is that, although
the lightest fermionic technibaryon might well be electromagnetically neutral and stable, 
its nominally unstable charged
partner might form stable charged techninuclei~\footnote{Much as the unstable neutron is
present in the Universe today in stable nuclei.}, whose abundance could exceed
the experimental upper limits on charged relics from the Big Bang~\cite{CEO3}.
Thus the rotational correction (\ref{MNcorrection}) to the techniSkyrmion mass is likely to be irrelevant.
Nevertheless, the semiclassical
mass estimates made here should be treated with caution.

\section{Relation to Vector-Meson Parameters}

Before closing, we recall that a connection may exist between the fourth-order Lagrangian
coefficients and the masses $M_{V_i}$ and couplings $G_{V_i}$ of vector mesons. In the framework of resonance chiral theory~\cite{EGPR}, 
one finds~\cite{Pichfit}
\beq
\alpha_4 \; = \; 2 \alpha_5 \; = \; \frac{G_V^2}{4 M_V^2}
\label{Pich}
\eeq
where in a single-meson dominance approximation it is estimated that $G_V = v/\sqrt{2}$.
These relations can be combined with the constraints (\ref{Conchabounds}, \ref{LHCsensitivity})
to derive present bounds on and prospective LHC sensitivities to $M_V$ in the single-resonance approximation.
In both cases, we find that the constraint on $\alpha_4$ is more powerful, yielding
\bea
{\rm Present~bound:} & M_V & > \; 350 \; {\rm GeV} , \nonumber \\
{\rm Future~sensitivity:} & M_V & > \; 700 \; {\rm GeV} .
\label{MV}
\eea
We recall that the corresponding relations in QCD are qualitatively successful
at the $\sim 30$\% level~\cite{Pichfit}. However, much as the QCD analogy may not
apply to the calculation of the $S$ parameter, the single-resonance
approximation may not be valid in the class of strongly-interacting electroweak models of interest~\cite{WTC}.

\section{Discussion}

In this paper we have explored the connection between the possible mass of the lightest
electroweak baryon (technibaryon) and fourth-order terms in the low-energy expansion
of the effective Lagrangian for electroweak symmetry breaking that is provided by the
Skyrme model. Fundamental considerations based on crossing, Lorentz invariance and
unitarity~\cite{DGPR} do not provide very stringent constraints on the possible electroweak baryon mass,
nor do present electroweak data~\cite{EGM}. However, a significant improvement in the indicative upper
limit could be provided by LHC data~\cite{EGM}, and further improvement could presumably be obtained
using the more precise measurements of $VV$ scattering that would become possible at a
high-energy $e^+ e^-$ collider.

It is an interesting question whether measurements of $VV$ scattering combined with other
experiments could eventually overconstrain or exclude a strongly-interacting electroweak
sector of the type discussed here. In this respect, an important issue is whether the lightest
electroweak baryon is stable, or not. We recall that anomalous electroweak interactions
could, in principle, cause the lightest electroweak baryon to decay~\cite{RST} into final states involving all
the electroweak doublets of the Standard Model, depending on the structure of the new
strongly-interacting sector. The lifetime of the lightest electroweak baryon could in principle be very short in some
models, if its mass exceeds a certain critical value~\cite{RST}. 

If the lightest electroweak baryon is stable on a cosmological time scale, dark matter scattering 
experiments have the potential to detect or exclude a massive neutral electroweak baryon~\cite{BDT,CEO3}. In
addition, bound states of neutral and charged electroweak baryons might form detectable, massive
`techninuclei'~\cite{CEO3}. If there 
is a more massive charged state with a lifetime exceeding a nanosecond, it could be detected in a
collider experiment. Electroweak baryons with shorter lifetimes could be detected through
their decays into a plethora of Standard Model particles.

The possible phenomenology of electroweak baryons is very rich, and depends on their
possible masses. As we have pointed out in this paper, there is a possible connection with
other electroweak observables that will surely repay further study.

\section*{Acknowledgements}

This study was triggered during the SCGT12 MiniWorkshop at the Kobayashi-Maskawa Institute
in Nagoya, and J.E. thanks K. Yamawaki and his colleagues for hospitality there.
We thank A. Pich for helpful correspondence. The work of J.E. was supported partly by the London
Centre for Terauniverse Studies (LCTS), using funding from the European
Research Council via the Advanced Investigator Grant 267352.

\end{document}